\documentstyle[aps,prc]{revtex}
\begin{document}
\draft
\title{Coulomb interaction between a spherical and a deformed nuclei}
\author{Noboru Takigawa,
\thanks{E-mail address: takigawa@nucl.phys.tohoku.ac.jp}
Tamanna Rumin
\thanks{E-mail address: rumin@nucl.phys.tohoku.ac.jp}
and Naoki Ihara 
\thanks{E-mail address: ihara@nucl.phys.tohoku.ac.jp}
}
\address{Department of Physics, Tohoku University, 
Sendai 980-8578, Japan}
\date{\today}

\maketitle

\begin{abstract}
We present analytic expressions of the Coulomb interaction between a 
spherical and a deformed nuclei which are valid for all separation 
distance. We demonstrate their significant deviations from commonly used 
formulae in the region inside the Coulomb radius, 
and show that they remove various shortcomings of the conventional 
formulae. 
\end{abstract}


\section{Introduction}

The Coulomb interaction between two extended objects is a fundamental 
quantity in many problems of physics. 
One example is a heavy-ion collision, where the standard procedure is to 
approximate the Coulomb interaction between the projectile and 
target nuclei by simply replacing the target radius by the sum of the 
projectile and target radii or by the so called Coulomb radius 
in the formula for the Coulomb interaction between a 
point charge and an extended target nucleus \cite{frob}. 
This approximation works well as long as the physically relevant region is 
outside the Coulomb radius. 
However, one definitely needs to improve the formulae if the region inside 
the Coulomb radius becomes to play an important role. 

In this paper, we present analytic expressions of the Coulomb interaction
between a spherical projectile and a deformed target which 
are valid for any separation distance between them 
and remove various shortcomings in the standard formulae. 
Since the formulae would have general values, here we mainly concentrate on  
derivation of the analytic expressions and 
comparing with commonly used formulae. 
The application to actual problems will be reported in separate papers.  

In sect.2 we explain the basic idea of the method. It gives 
various components of the Coulomb interaction such as the bare Coulomb 
interaction or inelastic scattering form factors in the linear or 
higher order couplings in terms of a one-dimensional Fourier integral. 
In sect.3, we present their analytic expressions obtained by 
computer assisted performance of the Fourier integrals. 
In sect.4 we compare the bare potential and the linear as well as 
the second order coupling form factors calculated by our formulae 
with those by commonly used formulae by taking the reaction of $^{16}$O with 
$^{238}$U as an example. 
We summarize the paper in sect.5, where we briefly mention some possible 
systems where the improved formulae will become important.  

\section{Coulomb interaction in the Fourier transform representation}

Denoting the densities of the projectile and target by 
$\rho_P$ and $\rho_T$, the Coulomb interaction between them is given by  
\begin{eqnarray}
V(\vec{R},\alpha) =
\int d\vec{r_1} \int d\vec{r_2} \frac{1}{\vert \vec{R}+\vec{r_2}
-\vec{r_1}\vert}\rho _P(\vec{r_2})\rho _T(\vec{r_1})
\label{ciorg}
\end{eqnarray}
where $\vec{R}$ is the position vector of the center of mass of the 
projectile measured from that of the target nucleus and 
describes their relative motion. $\alpha$ represents 
the ensemble of intrinsic coordinates, which are implicit in 
$\rho_P$ and $\rho_T$. They are the deformation parameters 
in the collective model which we adopt in the following. 
The key idea is to express the same quantity by the following Fourier 
transform representation \cite{Krappe,brog},
\begin{eqnarray}
V(\vec{R},\alpha) =
\frac{1}{2\pi^2}\int_0^\infty dk \int d\Omega_k \int d\vec{r_1} \int d\vec{r_2} \rho _P(\vec{r_2})\rho _T(\vec{r_1})e^{ik.(\vec{R}+\vec{r_2}-\vec{r_1})}
\label{cift}
\end{eqnarray}

For simplicity we assume a uniformly charged object with a sharp surface 
for both projectile and target, i.e.  
\begin{eqnarray}
\rho_i(\vec{r}) = \rho_i^{(0)} \Theta(R_i(\Omega)-r)
\label{dss}
\end{eqnarray}
where $\Theta(x)$ is a step function and 
the index i refers to either the projectile or the target. 
The angle dependent radius is given in terms of the deformation parameters as
\begin{eqnarray}
R_i(\Omega)=R_i^{(0)}\Big[1+\sum_{\lambda,\mu}\alpha_{\lambda,\mu}
Y^*_{\lambda,\mu}(\theta,\phi)\Big]
\label{radius}
\end{eqnarray}
The normalization condition then gives
\begin{eqnarray}
\rho_i^{(0)} = \frac{3Z_ie}{4\pi R_i^3}N_i(\alpha)
\label{rho0}
\end{eqnarray}
with
\begin{eqnarray}
N_i(\alpha) = \Big[1+\frac{3}{\sqrt{4\pi}}\alpha^{(i)}_{00}+
\frac{3}{4\pi}\sum_{\lambda \mu}\vert\alpha^{(i)}_{\lambda \mu}\vert^2+
\vartheta((\alpha^{(i)})^3)\Big]^{-1}
\label{norm}
\end{eqnarray}
One usually chooses $\alpha_{00}$ to conserve the volume. In that case, 
$N_i$ is one. 
In the following, we consider a spherical projectile and a 
deformed target. Accordingly, we remove the index i to distinguish 
the projectile or target from the deformation parameters 
and denote $R_P^{(0)}$ and $R_T^{(0)}$ simply by $R_P$ and $R_T$.
The integration over $\vec{r_2}$ in eq.(\ref{cift}) can be easily 
performed, and one obtains 
\begin{eqnarray}
V(\vec{R},\alpha) =
32\pi\rho^{(0)}_PR_P^3\sum_{\lambda \mu}Y^*_{\lambda \mu}(\Omega_R)
\int^\infty_0dk j_\lambda(kR)\frac{j_1(kR_P)}{kR_P}M^{(T)}_{\lambda \mu}(k)
\label{ciim1}
\end{eqnarray}
where 
\begin{eqnarray}
M^{(T)}_{\lambda \mu}(k)
&=&\int d{\vec{r}}j_\lambda(kr)Y_{\lambda \mu}(\Omega_r)\rho_T({\vec{r}})\label{mt1}\\
&=&\frac{\rho^{(0)}_T}{k^3}\int d\Omega_rY_{\lambda \mu}(\Omega_r)
\int^{kR(\Omega_r)}_0x^2j_\lambda(x)dx
\label{mt2}
\end{eqnarray}
By expanding with respect to the deformation parameters, one obtains
\begin{eqnarray}
M^{(T)}_{\lambda \mu}(k)
=\frac{\rho_T^{(0)}}{k^3}\Big[& &\delta_{\lambda 0}\sqrt{4\pi}
\int_0^{x}x^2 j_0(x)dx\nonumber\\
&+&x^3 j_{\lambda}(x)\alpha_{\lambda \mu}\nonumber\\
&+&x^2\big\{xj_{\lambda}(x)
+\frac{x^2}{2}\frac{dj_{\lambda}(x)}{dx}\big\}\alpha_{\lambda_1\mu_1}\alpha_{\lambda_2\mu_2}
\int d\Omega Y_{\lambda \mu}Y_{\lambda_1\mu_1}^{\star} Y_{\lambda_2\mu_2}^{\star}+...\Big]
\label{mtex}
\end{eqnarray}
where $x$  = $kR_T$. 
The first, second and third terms in the square brackets of eq.(\ref{mtex}) 
give the bare Coulomb interaction, the linear and the second order 
Coulomb couplings, respectively. We thus represent the total Coulomb 
interaction as
\begin{eqnarray}
V(\vec{R},\alpha)=V^{(0)} (R)+V^{(1)} (\vec{R},\alpha)+
V^{(2)} (\vec{R},\alpha)+......
\label{cisp}
\end{eqnarray}
where the first three terms are given by 
\begin{eqnarray}
&&V^{(0)} (R)=Z_PZ_Te^2F^{(0)}(R)  
\label{cimon}
\\
&&V^{(1)} (\vec{R},\alpha)=Z_PZ_Te^2
\sum_{\lambda,\mu}F^{(1)}_\lambda(R)Y^*_{\lambda \mu}(\Omega_R)\alpha_{\lambda \mu}
\label{cilin}
\\
&&V^{(2)} (\vec{R},\alpha)=Z_PZ_Te^2
\sum_{\lambda_1,\mu_1,\lambda_2,\mu_2}
\sum_{\lambda,\mu}F^{(2)}_\lambda(R)Y^*_{\lambda \mu}(\Omega_R)
\alpha_{\lambda_1,\mu_1}\alpha_{\lambda_2,\mu_2}
\int d\Omega Y_{\lambda \mu}Y_{\lambda_1\mu_1}^{\star} Y_{\lambda_2\mu_2}^{\star}
\label{cisec}
\end{eqnarray}
with the form factors defined by
\begin{eqnarray}
&&F^{(0)}(R)=\frac{18}{\pi}\int_0^{\infty} j_0(kR)
\frac{j_1(kR_P)}{kR_P}\frac{j_1(kR_T)}{kR_T} dk
\label{ffmon}
\\
&&F^{(1)}_\lambda(R)=\frac{18}{\pi}\int_0^{\infty} dk 
j_{\lambda}(kR)\frac{j_1(kR_P)}{kR_P} j_{\lambda}(kR_T)
\label{fflin}
\\
&&F^{(2)}_\lambda(R)=\frac{18}{\pi}\int_0^{\infty} 
j_{\lambda}(kR)\frac{j_1(kR_P)}{kR_P}\Big\{j_{\lambda}(kR_T)+
\frac{kR_T}{2}\frac{dj_{\lambda}(kR_T)}{d(kR_T)}\Big\}dk
\label{ffsec}
\end{eqnarray}
If we take the rotating frame where the z-axis is chosen to be parallel 
to the coordinate of the relative motion $\vec{R}$, as is 
often done in the studies of heavy ion fusion reactions, and 
if we assume an axially symmetric quadrupole deformation for the target 
nucleus, then the angular momentum algebra can be explicitly carried out and 
we obtain the following expressions for the linear and the 
leading second order Coulomb couplings 
\begin{eqnarray}
V^{(1)} (R,\beta_2,\theta)=Z_PZ_Te^2 
\sum_{\lambda=2,4,6}F^{(1)}_{\lambda}(R)\beta_{\lambda}Y_{\lambda0}(\theta,0)
\end{eqnarray}
\begin{eqnarray}
V^{(2)} (R,\beta_2,\theta)=Z_PZ_Te^2 
\Big[F^{(2)}_{\lambda=2}(R)\frac{\sqrt{5}}{7}\frac{1}{\sqrt{\pi}}
Y_{20}(\theta,0)
+F^{(2)}_{\lambda=4}(R)\frac{3}{7}\frac{1}{\sqrt{\pi}}Y_{40}(\theta,0)
\Big]\beta_2^2
\label{cisecl}
\end{eqnarray}
where $\beta_2$ is the quadrupole deformation parameter of 
the target nucleus, and $\theta$ the Euler angle to specify the orientation 
of its axially symmetric axis in the rotating frame.

The important achievement of the Fourier transform method 
is that one needs to perform only 
one dimensional integral, whose results we will present 
in the next section. Before moving, we wish to 
comment that one can easily extend the same procedure to incorporate 
the surface diffuseness of the colliding nuclei by smearing the sharp 
surface in eq.(\ref{dss}) with a Yukawa function \cite{Krappe,brog}. 
The change is simply that the integrand of the Fourier transform 
representation of each 
form factor eqs.(\ref{ffmon}) through (\ref{ffsec}) gets two additional 
factors corresponding to the Fourier transforms of the 
Yukawa function which specify the surface properties 
of the projectile and target nuclei.

\section{Analytic expressions of the bare 
Coulomb potential and coupling form factors}

The Fourier integrals on the r.h.s. of eqs.(\ref{ffmon}) through (\ref{ffsec}) 
are tedious but still doable analytically by hands, 
since the integrands are given only by polynomials and trigonometric 
functions. 
However, computer assisted analytic integration is much easier and practical.
Introducing the parameters $R_C$ and $R_{CC}$ by 
$R_C=R_T+R_P$ and $R_{CC}=\vert(R_T-R_P)\vert$, the results read,

\medskip

\noindent
(1) bare Coulomb interaction, 
\begin{small}
\begin{eqnarray}
V^{(0)}(R)=Z_PZ_Te^2 \left\{ \begin{array}{ll}
1/R, &(R>R_C)
\vspace{0.3cm}\\
\Big[\{-R^5+24(R_P+R_T)^3(-R_P^2+3R_PR_T-R_T^2)
+45R(-R_P^2+R_T^2)^2\\
+15R^3(R_P^2+R_T^2)-40R^2(R_P^3+R_T^3)\}/(160R_P^3R_T^3)\\
+(R_P^6-9R_P^4R_T^2
+16R_P^3R_T^3-9R_P^2R_T^4+R_T^6)/(32RR_P^3R_T^3)\Big], &(R_C>R>R_{CC})
\vspace{0.3cm}\\
\{-5R^2+3(-R_P^2+5R_T^2)\}/(10R_T^3), &(R_{CC}>R>0)\\
\end{array} \right.
\label{cifin}
\end{eqnarray}
\end{small}
These formulae agree with those derived in \cite{Rowley} and used  in 
\cite{baha}. 

\medskip

\noindent
(2) quadrupole Coulomb coupling form factor of linear order,
\begin{footnotesize}
\begin{eqnarray}
F^{(1)}_{\lambda=2}(R)=\left\{ \begin{array}{ll}
3 R_T^2/(5 R^3), &(R>R_C)
\vspace{0.3cm}\\
\Big[3R^2/(10R_T^3)+\{3R^5-12R^3(3R_P^2+R_T^2)+18R(-3R_P^4+2R_P^2R_T^2
+R_T^4)\}/(256R_P^3R_T^3)\\
+3(R_P^6-3R_P^4R_T^2+3R_P^2R_T^4-R_T^6)/(64RR_P^3R_T^3)\\
+3(-3R_P^8+20R_P^6R_T^2-90R_P^4R_T^4 +128R_P^3R_T^5-60R_P^2R_T^6+5R_T^8)/(1280R^3R_P^3R_T^3)\Big], &(R_C>R>R_{CC})
\vspace{0.3cm}\\
3R^2/(5R_T^3), &(R_{CC}>R>0)\\
\end{array} \right.
\label{ffl2fin}
\end{eqnarray}
\end{footnotesize}

\medskip

\noindent
(3) hexadecapole Coulomb coupling form factor of linear order,
\begin{footnotesize}
\begin{eqnarray}
F^{(1)}_{\lambda=4}(R)=\left\{ \begin{array}{ll}
R_T^4/(3 R^5),  & (R>R_C)
\vspace{0.3cm}\\
\Big[R^4/(6R_T^5)+\{7R^7-18R^5(7R_P^2+R_T^2)+9R^3(-35R_P^4+10R_P^2R_T^2
+R_T^4)\\
+4R(35R_P^6-45R_P^4R_T^2+9R_P^2R_T^4+R_T^6)\}/(2048R_P^3R_T^5)\\
+9(-7R_P^8+20R_P^6R_T^2-18R_P^4R_T^4 +4R_P^2R_T^6+R_T^8)/(2048RR_P^3R_T^5)\\
+9(R_P^{10}-5R_P^8R_T^2+10R_P^6R_T^4-10R_P^4R_T^6+5R_P^2R_T^8-R_T^{10})/(1024R^3R_P^3R_T^5)\\
+(-7 R_P^{12}+54 R_P^{10} R_T^2-189 R_P^8 R_T^4 +420 R_P^6 R_T^6-945 R_P^4 R_T^8+1024 R_P^3 R_T^9\\
-378 R_P^2 R_T^{10}+21 R_T^{12})/(6144 R^5 R_P^3 R_T^5)\Big],    
& (R_C>R>R_{CC})
\vspace{0.3cm}\\
R^4/(3R_T^5), & (R_{CC}>R>0)\\
\end{array} \right.
\label{ffl4fin}
\end{eqnarray}
\end{footnotesize}

\medskip

\noindent
(4) hexacontatetrapole Coulomb coupling form factor of linear order,
\begin{scriptsize}
\begin{eqnarray}
F^{(1)}_{\lambda=6}(R)=\left\{ \begin{array}{ll}
3 R_T^6/(13 R^7), & (R>R_C)
\vspace{0.3cm}\\
\Big[3 R^6/(26 R_T^7)+\{99 R^9-216 R^7 (11 R_P^2+R_T^2)
+84 R^5 (-99 R_P^4+18 R_P^2 R_T^2+R_T^4)\\
+24 R^3 (231 R_P^6-189 R_P^4 R_T^2+21 R_P^2 R_T^4+R_T^6)
+18 R (-231 R_P^8+420 R_P^6 R_T^2 \\
-210 R_P^4 R_T^4+20 R_P^2 R_T^6+R_T^8)\}/(65536 R_P^3 R_T^7)\\
+3 (99 R_P^{10}-315 R_P^8 R_T^2+350 R_P^6 R_T^4-150 R_P^4 R_T^6
+15 R_P^2 R_T^8+R_T^{10})/(8192 R R_P^3 R_T^7)\\
+21 (-11 R_P^{12}+54 R_P^{10} R_T^2-105 R_P^8 R_T^4
+100 R_P^6 R_T^6-45 R_P^4 R_T^8+6 R_P^2 R_T^{10}
+R_T^{12})/(16384 R^3 R_P^3 R_T^7)\\
+27 (R_P^{14}-7 R_P^{12} R_T^2+21 R_P^{10} R_T^4-35 R_P^8 R_T^6+35 R_P^6 R_T^8
-21 R_P^4 R_T^{10}+7 R_P^2 R_T^{12}-R_T^{14})/(8192 R^5 R_P^3 R_T^7)\\
+3 (-99 R_P^{16}+936 R_P^{14} R_T^2-4004 R_P^{12} R_T^4+10296 R_P^{10} R_T^6
-18018 R_P^8 R_T^8+24024 R_P^6 R_T^{10}\\
-36036 R_P^4 R_T^{12}+32768 R_P^3 R_T^{13}-10296 R_P^2 R_T^{14}
+429 R_T^{16})/(851968 R^7 R_P^3 R_T^7)\Big], & (R_C>R>R_{CC})
\vspace{0.3cm}\\
3R^6/(13R_T^7), &(R_{CC}>R>0)
\end{array} \right.
\label{ffl6fin}
\end{eqnarray}
\end{scriptsize}

\medskip

\noindent
(5) quadrupole Coulomb coupling form factor of second order,
\begin{scriptsize}
\begin{eqnarray}
F^{(2)}_{\lambda=2}(R)=\left\{ \begin{array}{ll}
6 R_T^2/(5 R^3), & (R>R_C)
\vspace{0.3cm}\\
\Big[-3R^2/(20R_T^3)-\{3R^5-12R^3(3R_P^2-R_T^2)-18R(3R_P^4+2R_P^2R_T^2
+3R_T^4)\}/(512R_P^3R_T^3)\\
+3(-R_P^6-3R_P^4R_T^2+9R_P^2R_T^4-5R_T^6)/(128RR_P^3R_T^3)\\
+3(3R_P^8+20R_P^6R_T^2-270R_P^4R_T^4+512R_P^3R_T^5-300R_P^2R_T^6+35R_T^8)/(2560R^3R_P^3R_T^3)\Big], & (R_C>R>R_{CC})
\vspace{0.3cm}\\
-3R^2/(10R_T^3), & (R_{CC}>R>0)
\end{array} \right.
\label{ffs2fin}
\end{eqnarray}
\end{scriptsize}

\medskip

\noindent
(6) hexadecapole Coulomb coupling form factor of second order,
\begin{scriptsize}
\begin{eqnarray}
F^{(2)}_{\lambda=4}(R)=\left\{ \begin{array}{ll} 
R_T^4/(R^5), &(R>R_C)
\vspace{0.3cm}\\
\Big[-R^4/(4R_T^5)- \{21R^7-18R^5(21R_P^2+R_T^2)-9R^3(105R_P^4-10R_P^2R_T^2
+R_T^4)\\
-12R(-35R_P^6+15R_P^4R_T^2+3R_P^2R_T^4+R_T^6)\}/(4096R_P^3R_T^5)\\
+9(21R_P^8-20R_P^6R_T^2-18R_P^4R_T^4+12R_P^2R_T^6+5R_T^8)/(4096RR_P^3R_T^5)\\
+9(-3R_P^{10}+5R_P^8R_T^2+10R_P^6R_T^4-30R_P^4R_T^6+25R_P^2R_T^8
-7R_T^{10})/(2048R^3R_P^3R_T^5)\\
+(7R_P^{12}-18R_P^{10}R_T^2-63R_P^8R_T^4+420R_P^6R_T^6
-1575R_P^4R_T^8+2048R_P^3R_T^9-882R_P^2R_T^{10}\\
+63R_T^{12})/(4096R^5R_P^3R_T^5) \Big], & (R_C>R>R_{CC})
\vspace{0.3cm}\\
-R^4/(2R_T^5), & (R_{CC}>R>0)
\end{array} \right.
\label{ffs4fin}
\end{eqnarray}
\end{scriptsize}

\section{Comparison with commonly used formulae}

We now compare our formulae with three commonly used models, which are given by

\medskip

\noindent
(A) model I (point projectile model) 
\begin{eqnarray}
&&V^{(0)}(R)=\frac{Z_PZ_Te^2}{R}
\label{pcoul}
\\
&&F^{(1)}_\lambda(R)=\frac{3}{2\lambda+1}\frac{R_T^\lambda}{R^{\lambda+1}}
\label{pcff1}
\\
&&F^{(2)}_\lambda(R)=\frac{6}{2\lambda+1}\frac{R_T^\lambda}{R^{\lambda+1}}
\label{pcff2}
\end{eqnarray}

\medskip

\noindent
(B) model II (uniform charge model 1)
\begin{eqnarray}
&&V^{(0)}(R)= Z_PZ_Te^2 \left\{ \begin{array}{ll}
1/R, & (R>R_C)
\\  
1/(2R_C)[3-(\frac{R}{R_C})^2], & (R<R_C)\\
\end{array} \right.
\label{uacoul}
\\
&&F^{(1)}_\lambda(R)=\widetilde{\beta_{\lambda}^c}\frac{3}{(2\lambda+1)}
\left\{ \begin{array}{ll}
R_C^{\lambda}/R^{\lambda+1}, & (R>R_C)
\\  
R^{\lambda}/(R_C)^{\lambda+1}, & (R<R_C)
\label{uaff}
\\
\end{array} \right.
\end{eqnarray} 
with 
\begin{eqnarray}
\widetilde{\beta_{\lambda}^c}=\frac{R_T^{\lambda}\beta_{\lambda}^c}
{R_C^{\lambda}}
\label{scale}
\end{eqnarray}

\medskip

\noindent
(C) model III (uniform charge model 2)
\begin{eqnarray}
&&V^{(0)}(R)= Z_PZ_Te^2 \left\{ \begin{array}{ll}
1/R, & (R>R_T)
\\  
1/(2R_T)[3-(\frac{R}{R_T})^2], & (R<R_T)
\label{ubcoul}
\\
\end{array} \right.
\\
&&F^{(1)}_\lambda(R)=\frac{3}{(2\lambda+1)}
\left\{ \begin{array}{ll}
R_T^{\lambda}/R^{\lambda+1}, & (R>R_T)
\\  
R^{\lambda}/(R_T)^{\lambda+1}, & (R<R_T)
\label{ubff}
\\
\end{array} \right.
\end{eqnarray} 
The model I is used in almost all analyses 
of heavy-ion fusion reactions at energies near and below the 
Coulomb barrier. The model II can be often found in textbooks 
of heavy-ion collisions \cite{frob}.
Model I and II can be too crude concerning the bare potential. 
The model II and III have a shortcoming that each coupling form 
factor has a cusp at the separation distance, where the asymptotic formulae 
are matched to the formulae in the short distance region.  
The derivative of the coupling form factor is 
discontinuous at that distance. 
In writing eqs.(\ref{uaff}) and (\ref{scale}), we have 
assumed that the target nucleus has an axially symmetric static deformation 
with the Coulomb deformation parameter $\beta_{\lambda}^c$.  
Eq.(\ref{scale}) is the scaling condition of the deformation parameter 
to guarantee the correct coupling in the asymptotic region. 

Figs.1 through 3 compare the bare Coulomb potential and the 
linear and quadratic coupling form factors $F^{(1)}_{\lambda=2,4,6}(R)$ 
and $F^{(2)}_{\lambda=2,4}(R)$ calculated by these formulae and 
by our improved formulae. The form factors have been multiplied with 
$Z_PZ_Te^2$ to make the ordinate of all figures have the dimension of energy. 
They have been calculated for the scattering of 
$^{16}$O with $^{238}$U. The Coulomb radius parameter has been chosen to be 
1.06 fm. These figures clearly show the shortcomings of all the 
three commonly used simple models, which are solved by our new formulae. 

The important question is whether the deviations of the conventional 
models from our improved formulae play some significant roles 
in actual problems of physics, e.g. in analyzing heavy-ion fusion reactions 
at energies near and below the Coulomb barrier.  
In order to have some idea on this question, we again consider 
the fusion reaction of a spherical projectile $^{16}$O with 
an axially symmetric deformed target $^{238}$U. 
It is now well known that the excitation of the 
ground state rotational band of $^{238}$U plays an important role in 
enhancing the fusion cross section in this reaction.
Instead of performing full coupled-channels calculations 
to take this effect into account, 
one often describes this reaction based on the no-Coriolis and 
the sudden tunneling, i.e. degenerate spectrum limit, approximations. 
In this case, the fusion probability for each partial wave J 
is given by first calculating the fusion probability $P_J(E,\theta)$ for a 
fixed orientation of the target nucleus $\theta$ 
and then taking average over $\theta$
\begin{eqnarray}
P_J(E)=\frac{1}{2}\int_0^{\pi}P_J(E,\theta)\sin\theta d\theta, 
\label{pene}
\end{eqnarray}
The effective potential for each orientation $\theta$ is given by
\begin{eqnarray}
V_J(R,\beta_2,\theta)=V_N(R,\beta_2,\theta)+
\frac{\hbar^2}{2 \mu R^2}J(J+1)+Z_PZ_Te^2
(F^{(0)}(R)+\beta_2 F^{(1)}_{\lambda=2}(R) Y_{20}(\theta,0))
\label{effpot}
\end{eqnarray}
where $V_N(R,\beta_2,\theta)$ is the nuclear potential and $\mu$ the reduced 
mass between the projectile and target. 
Since our aim is not to perform quantitative analyses, but 
to illustrate under what circumstances our improved formulae 
show its power, we took a simple model which considers only 
quadrupole deformation $\beta_2=0.289$ \cite{tama} for $^{238}$U and linear 
coupling. The Gauss integral in eq.(\ref{pene}) is replaced by the 
$\frac{J_{max}}{2}$+1 points Gauss 
quadrature if the rotational excitation is truncated at $J_{max}$
\begin{eqnarray}
P_J(E)=\frac{1}{2}\sum_iw_iP_J(E,\theta_i), 
\label{penefn}
\end{eqnarray}
where $\theta_i$ are the angles, where the Legendre polynomial 
$P_{J_{max}+2}(\theta)$ becomes zero. 

The active angles are the zeros of $P_4(cos\theta)$,which are about 
30.55 and 70.12 degrees, if $J_{max}$=2.
In Fig.4 we show the effective s-wave potentials for these two angles 
calculated by our improved formulae (the solid line) 
and by three conventional models (the dotted, dashed and dot-dashed lines). 
We assumed a Wood-Saxon potential for the nuclear potential, 
where the radius parameter has been chosen such that 
$R_N=R_N^{(0)}+R_T \beta_{2}Y_{20}(\theta,0)$ with 
$R_N^{(0)}=R_P+R_T=1.06(A_P^{1/3}+A_T^{1/3})$.

We see clear differences among four models. 
However, these differences do not have any physical significance 
for $^{16}$O+$^{238}$U fusion reactions at low energies, 
since the deviation is localized well inside the barrier region, while 
the fusion probability is governed by the barrier property, which 
is the same for all four calculations. Similar situation 
will hold in general for medium weight heavy-ion collisions.  
This is a natural consequence of 
taking the same values for the nuclear and Coulomb radii. 
We can, however, think of some cases where the difference 
plays an important role as will be mentioned below. 

\section{Summary and discussions}

We have derived analytic expressions of the Coulomb interaction, 
which are valid for any separation distance between the 
projectile and target, and have demonstrated their significant deviations 
from commonly used models. 
Our new formulae solve the 
cusp and discontinuity problems in the form factors and 
in their derivatives in commonly used models.  

We argued that these deviations will not cause any 
significant effects on the fusion reactions between two 
heavy nuclei such as the $^{16}$O+$^{238}$U fusion reactions 
at energies near and below the Coulomb barrier, 
which have been very popular subjects of nuclear physics 
in the past decades. 
However, we have to keep in mind that this conclusion has 
been drawn 
by assuming the same values for the nuclear and Coulomb radii. 
It is then natural that the deviation of the 
conventional models from our improved formulae takes place 
well inside the barrier region. 
One interesting system will therefore be the system where the Coulomb 
radius is larger than the nuclear radius.  
One such example could be heavy-ion collisions induced by 
unstable neutron deficient isotopes. 
Another interesting case will be the fusion as well as 
elastic and inelastic scatterings between light heavy ions, 
where the absorption in the internal region is not so strong. 
In this case, the differences among 4 models at short distances 
will lead to quite different cross sections from each other. 
 
\section*{Acknowledgments}

N.T. wishes to thank Dr. Krappe for useful discussions, and 
Prof. J. Eichler and his colleagues at the Hahn Meiner Institute, 
Berlin, for many discussions and kind hospitality. 


\newpage
\begin{center}
{\Large Figure Captions}
\end{center}

\noindent
{\large Fig.1}\\
Comparison of bare Coulomb interaction. 
The solid, dotted, dashed and dot-dashed lines have been calculated 
based on our improved formulae, Model I through III, respectively. 

\noindent
{\large Fig.2}\\
Comparison of the linear Coulomb coupling form factor plotted in energy scale 
(see text). Notation is the same as in Fig.1. (a) quadrupole coupling 
 (b) hexadecapole coupling (c) hexacontatetrapole coupling. \\

\noindent
{\large Fig.3}\\
Comparison of the second order 
Coulomb coupling form factor plotted in energy scale. 
The solid and the dotted lines are for our improved formulae and 
Model I. \\

\noindent
{\large Fig.4}\\
Comparison of the effective s-wave barrier calculated 
based on eq.(\ref{effpot}) with our improved formulae and 3 conventional 
models. The results are shown for two orientations, 30.6 and 70.1 degrees. 
The lower and the upper barriers correspond to the former and the latter,
respectively. 

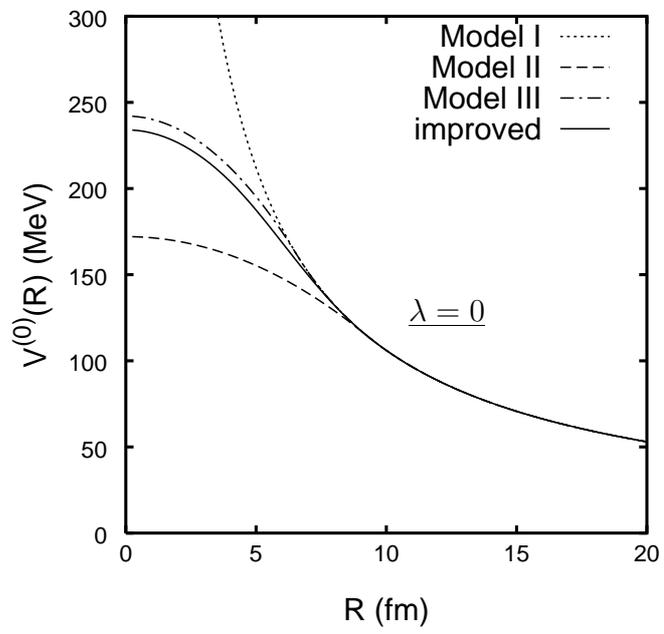
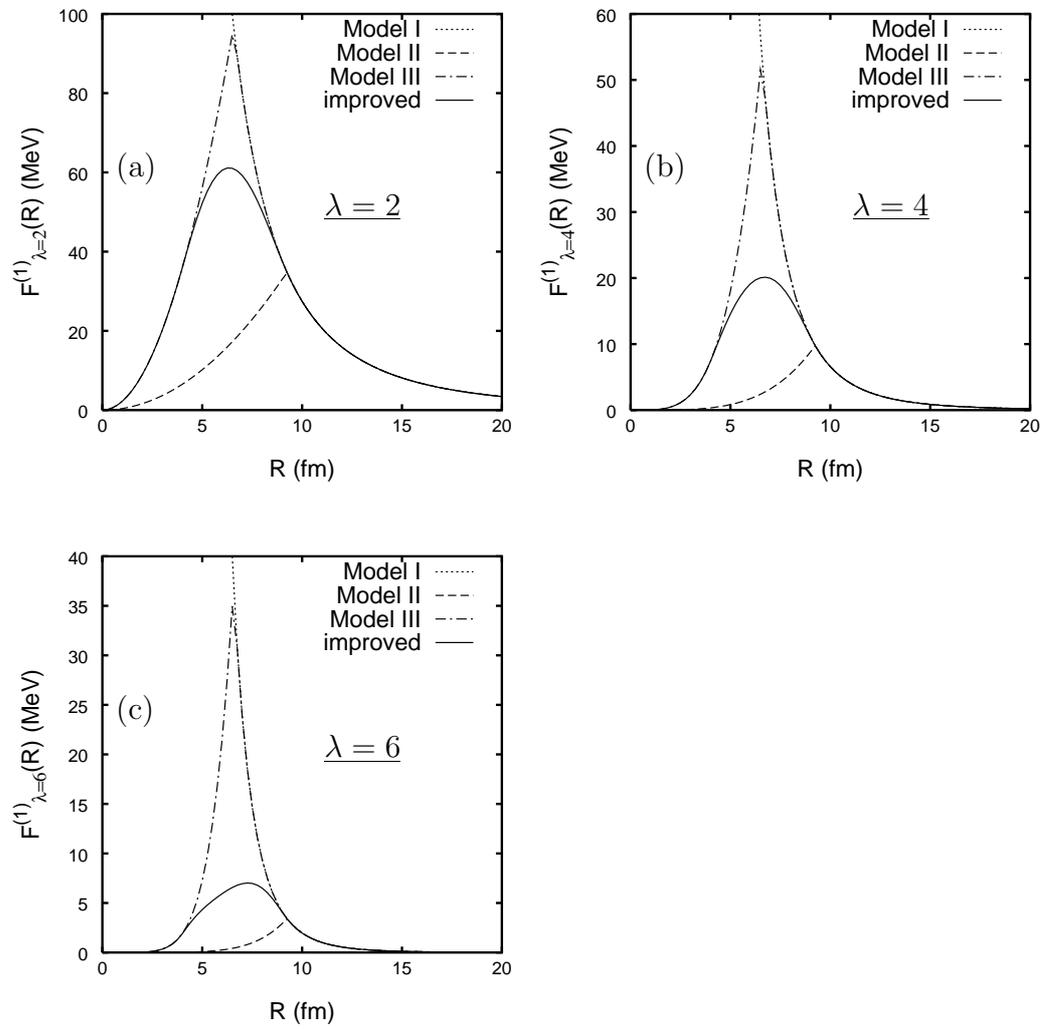
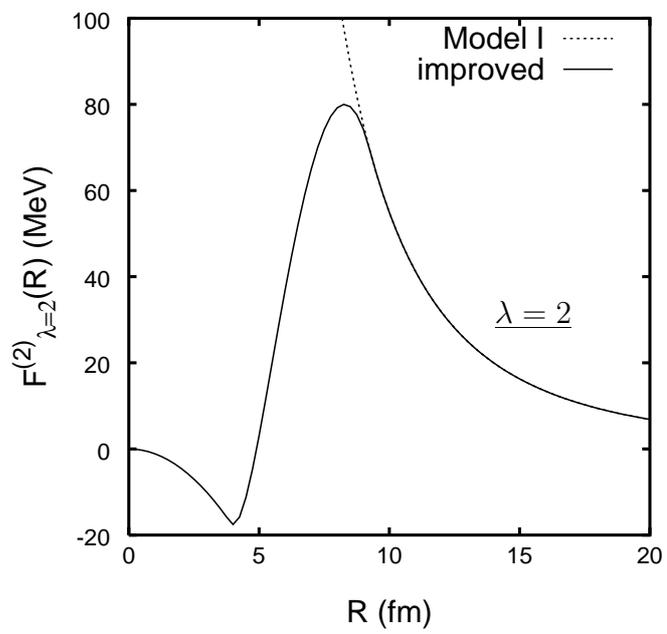
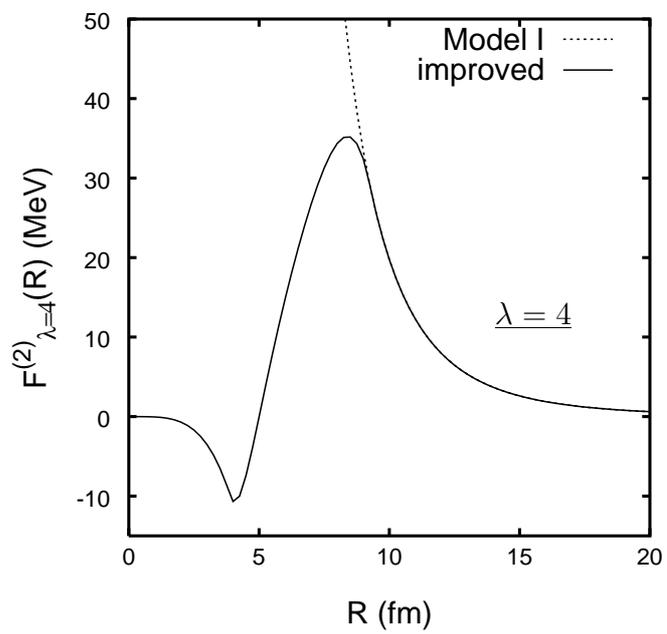
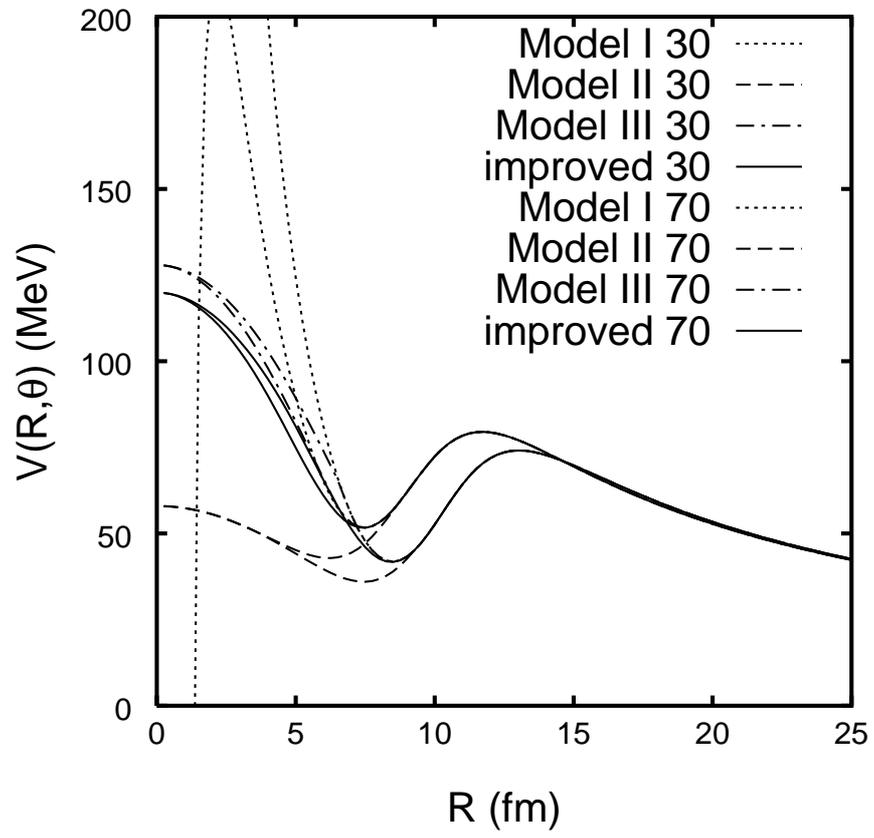
\end{document}